\documentclass[aps,prd,nofootinbib,twocolumn,superscriptaddress,floatfix]{revtex4}
\usepackage{mathrsfs}
\usepackage{latexsym}
\usepackage{amsmath}
\usepackage{amssymb}
\usepackage{graphicx}
\usepackage{longtable}
\usepackage{multirow}
\usepackage{bbm}
\usepackage{epsfig}
\usepackage[usenames]{color}
\usepackage[colorlinks=true,citecolor=blue,linkcolor=blue]{hyperref}
\allowdisplaybreaks

\begin{document}



\title{Speed of sound in QCD matter}
\author{Wei-bo He} 
\affiliation{ MOE Key Laboratory for Non-equilibrium Synthesis and Modulation of Condensed Matter, School of Physics, Xi’an Jiaotong University, Xi’an 710049, China}
\affiliation{ School of Physics, Peking University, Beijing, 100871, China}

\author{Guo-yun Shao}   
\email[Corresponding author: ]{gyshao@mail.xjtu.edu.cn} 
\affiliation{ MOE Key Laboratory for Non-equilibrium Synthesis and Modulation of Condensed Matter, School of Physics, Xi’an Jiaotong University, Xi’an 710049, China}

\author{Xue-yan Gao}
\affiliation{ MOE Key Laboratory for Non-equilibrium Synthesis and Modulation of Condensed Matter, School of Physics, Xi’an Jiaotong University, Xi’an 710049, China}

\author{Xin-ran Yang}   
\affiliation{ MOE Key Laboratory for Non-equilibrium Synthesis and Modulation of Condensed Matter, School of Physics, Xi’an Jiaotong University, Xi’an 710049, China}

\author{Chong-long Xie}
\affiliation{ MOE Key Laboratory for Non-equilibrium Synthesis and Modulation of Condensed Matter, School of Physics, Xi’an Jiaotong University, Xi’an 710049, China}

\begin{abstract}
We systematically investigate the speed of sound in QCD matter under different conditions in the grand canonical ensemble within the Polyakov loop improved Nambu--Jona-Lasinio (PNJL)  model. The numerical results indicate that the dependence of speed of sound  on   parameters like temperature and chemical potential can be indicative of QCD phase transition.  
Some new features of speed of sound are  discovered, for instance,  the hierarchy of sound velocity for $u(d)$ and $s$ quark at low temperature with the increasing  chemical potential and the squared sound velocity approaching to almost zero in the critical region. 
We  also formulate the relations between differently defined sound velocity using the fundamental thermodynamic relations. Some conclusions derived are useful for  hydrodynamics simulation and calculation of transport coefficient of bulk viscosity.
%
%
%
\end{abstract}


\maketitle
\section{introduction}

Heavy-ion collisions at relativistic energies can create strongly interacting hot and/or dense matter.
The exploration of QCD phase structure and  search for phase transition signatures are significant goals in both theory and heavy-ion collision experiments.
The calculations from  first principle  lattice QCD (LQCD) indicate that the transformation from quark-gluon plasma~(QGP) to hadrons  is a smooth crossover~\cite{Aoki06, Gupta11, Bazavov12, Borsanyi13,Bazavov14, Bazavov17,Borsanyi14,Borsanyi20}  at high temperature and small baryon chemical potential.
A first-order phase transition, with a critical endpoint (CEP) connecting with a crossover  transformation, is predicted at large chemical potential by  quark models which share the symmetry properties of QCD,  Dyson-Schwinger equation (DSE)  and functional renormalization group~(FRG) approaches ~(e.g.,\cite{Fukushima04,Ratti06,Costa10,Fu08, Sasaki12, Ferreira14, Schaefer10, Skokov11, Qin11,Gao16, Fischer14, Fu20}).

Some possible phase transition signals for such a phase structure were proposed based on the ratios of net-baryon number cumulants~\cite{Stephanov}. The cumulants of net proton~(proxy for baryon) have been measured in the Beam Energy Scan (BES-I) experiments at RHIC STAR, and a non-monotonic energy dependence of  the net-proton number kurtosis  $\kappa \sigma^2$ was discovered~\cite{Luo2014, Luo2016, Luo2017,Adam21}.  It  possibly hints that the STAR experiments with smaller colliding energy pass through the QCD critical region. More accurate measurement on BES-II and the relevant experimental projects at NICA/FAIR/J-PARC/HIAF in the near future will provide us more information about the QCD phase diagram.

The hydrodynamics simulation provides another scheme to explore the QCD phase transition. 
The space-time evolution of QCD matter  can be successfully described using the relativistic dissipative hydrodynamics~\cite{Song11, Song112, Deb16}. One of the most important quantities in hydrodynamics is the speed of sound~$c_s$  and its dependence on environment  (temperature, density, chemical potential, etc.)~\cite{Albright06}.
In heavy-ion collision experiments, the sound velocity  is a thermodynamic observable and carries important information in describing the evolution of the fireball. The study in \cite{Gardim20,Sahu21,Biswas20} shows that $c_s^2$ as a function of charged particle multiplicity $\langle d N_{ch}/d\eta \rangle$  can reveal the dynamics of heavy-ion collision. Recently, in \cite{Sorensen21} the authors  estimate $c_s$ as a function of the logarithmic derivative with respect to the baryon  density of QCD matter, and try to build a connection with the baryon number cumulants to aid in detecting the QCD critical endpoint. 

Besides heavy-ion collision experiments, the speed of sound is of particular interest to neutron star research~(e.g.,\cite{Reed20, Kanakis20,Han20}).
The behavior of $c_s$ as a function of density influences the mass-radius relation, the tidal deformability, and provides a sensitive probe of the equation of state~(EOS) of neutron star matter. To obtain neutron stars with masses above two solar masses, several studies find that it is essential for the nuclear EOS to have a region in which the EOS is very stiff, where the sound velocity square is significantly larger than $1/3$~\cite{Tews18,Greif19,Forbes19,Drischler20,Essick20,Han19,Kojo}. The study in \cite{Jaikumar21} also indicates that the speed of sound is crucial for the gravitational wave frequencies induced by the $g$-mode oscillation of a neutron star. 

It is also an interesting topic to study $c_s$ during the QCD phase transition in the early universe by observing the induced gravitational wave. Although the propagation of gravitational wave is insensitive to $c_s$, the sound speed value affects the dynamics of primordial density perturbations, and the induced gravitational waves by their horizon reentry can then be an indirect probe on both the EOS and sound velocity, which can provide useful knowledge of the evolution in the era of QCD phase transition~\cite{Abe21}. 

It can be seen from the above the speed of sound is a fundamental property of strongly interacting matter. Some calculations of sound velocity in QCD matter have been performed  in LQCD~\cite{Aoki06, Borsanyi14, Bazavov14, Philipsen13, Borsanyi20},  (P)NJL model~\cite{Motta18,Ghosh06,Marty13,Deb16,Saha18,zhao20}, quark-meson coupling model~\cite{Schaefer10,Abhishek18},  hadron resonance gas (HRG) model~\cite{Venugopalan92,Bluhm14}, field correlator method~(FCM)~\cite{Khaidukov18, Khaidukov19} and quasiparticle model~\cite{Mykhaylova21}.  
By far,  the main focus is put on the region of high temperature and  zero or small chemical potential. There is a crucial lack of  $c_s$ in the full phase diagram and its relation with QCD phase transitions, including  both the chiral and deconfinement  transition.
On the other hand, different definitions or even different statistical ensembles are taken to calculate the speed of sound in literature~(e.g., \cite{ Borsanyi20,Ghosh06,Marty13,Deb16,Saha18,zhao20,Abhishek18}). Some issues with respect to thermodynamic relations need to be clarified.

In this work, we give an intensive study on the sound velocity in QCD matter in the full phase diagram with different definitions of $c_s$ in the grand canonical ensemble. The numerical results indicate that the dependence of $c_s$ on the system parameters like temperature and/or chemical potential can be indicative of QCD phase transition.  
Some new features of speed of sound are also discovered for the first time, for example, the hierarchy phenomenon of sound velocity for $u(d)$ and $s$ quark at low temperature with the increasing  chemical potential and the squared sound velocity at the CEP  approaching to almost zero. 
We  also formulate the relations between differently defined $c_s$ using the fundamental thermodynamic relations, and derive some conclusions useful for hydrodynamics simulation.

The paper is organized as follows. In Sec.~II, we introduce the thermodynamic relations about the speed of sound for a thermal system, and then give a brief introduction to  the 2+1 flavor PNJL quark model. In Sec.~III, we show the numerical results of speed of sound, and discuss the relations under different definitions, as well as the indicative signals of QCD phase transition. A summary is finally given in Sec. IV.

\section{speed of sound and the PNJL quark model}
The speed of sound is a fundamental properties of any substance. The definition of speed of sound requires specifying a quantity constant during the propagation of the compression wave through a medium.
The square of speed of sound is usually defined as
\begin{equation}
\label{ }
c_x^2=\bigg(\frac{\partial p}{\partial \epsilon}\bigg)_x,
\end{equation}
where $p$ and $\epsilon$ are pressure and energy density, respectively, and $x$ denotes the parameter fixed in the calculation of the sound velocity.

For a fireball created in relativistic heavy-ion collisions, it evolves with constant entropy density per baryon $s/\rho_B$ in the ideal fluid, where $s$ is the entropy density and $\rho_B$ is the baryon number density. This conclusion can be derived in hydrodynamics due to the conservations of energy and baryon number, therefore it is meaningful to calculate the sound velocity along the isentropic curve
\begin{equation}
\label{ }
c_{s/{\rho_B}}^2=\bigg(\frac{\partial p}{\partial \epsilon}\bigg)_{s/{\rho_B}}.
\end{equation}
The dependence of $c_{s/{\rho_B}}^2$ on parameter like temperature and chemical potential can indicate the change of sound velocity during the evolution and provide important information of  interaction and  the equation of state of medium.

There are also two other definitions of speed of sound with constant entropy density or baryon number density usually used in the intermediate process of hydrodynamic evolution~\cite{Deb16},
\begin{equation}
\label{ }
c_s^2=\bigg(\frac{\partial p}{\partial \epsilon}\bigg)_s \,\,\,\text{and}\,\,\,\,c_{\rho_B}^2=\bigg(\frac{\partial p}{\partial \epsilon}\bigg)_{\rho_B}.
\end{equation}
To calculate the square of speed of sound under different definitions, it is necessary to derive the corresponding formulae as functions of $T$ and $\mu_B$ from the fundamental thermodynamic relations. We give here the  formulae derived in the grand canonical ensemble. First of all, the entropy density and baryon number density can be simply derived with 
\begin{equation}
s=-\bigg(\frac{\partial p}{\partial T}\bigg)_{\mu_B}\,\,\,\,\text{and}\,\,\,\rho_B=-\bigg(\frac{\partial p}{\partial \mu_B}\bigg)_T.
\end{equation}

The $c_x^2$ for different parameter $x$ can be derived using Jacobian methods and thermodynamic relations in the grand canonical ensemble as
\begin{equation}\label{crho}
c_{\rho_B}^{2}=\frac{\partial (p,\rho_B)}{\partial(\epsilon ,\rho_B)}=\frac{s \chi_{\mu \mu}-\rho_{B} \chi_{\mu T}}{T\left(\chi_{T T} \chi_{\mu \mu}-\chi_{\mu T}^{2}\right)} ,
\end{equation}
\begin{equation}\label{cs}
c_{s}^{2}=\frac{\partial (p,s)}{\partial(\epsilon ,s)}=\frac{\rho_{B} \chi_{T T}-s \chi_{\mu T}}{\mu_{B}\left(\chi_{T T} \chi_{\mu \mu}-\chi_{\mu T}^{2}\right)} ,
\end{equation}
and
\begin{equation}
c_{s/\rho_B}^{2}=\frac{\partial (p,s/\rho_B)}{\partial(\epsilon ,s/\rho_B)}=\frac{c_{\rho_B}^{2} T s+c_{s}^{2} \mu_B \rho_{B}}{Ts+\mu_B \rho_B}.
\end{equation}
In the above equations, the second-order susceptibility $\chi_{x,y}$ is defined as  $\chi_{x,y}=\partial^2 p/\partial x\partial y$.

Besides, it is also convenient to calculate the square of the speed of sound with a constant temperature or  chemical potential with 
\begin{equation}
\label{ }
c_T^2=\bigg(\frac{\partial p}{\partial \epsilon}\bigg)_T, \,\,\,\,\,\,\,\,c_{\mu_B}^2=\bigg(\frac{\partial p}{\partial \epsilon}\bigg)_{\mu_B},
\end{equation}

In the calculation, we take the popular 2+1 flavor PNJL quark model. 
The Lagrangian density in this model is given by
\begin{eqnarray}
\mathcal{L}&\!=&\!\bar{q}(i\gamma^{\mu}D_{\mu}\!+\!\gamma_0\hat{\mu}\!-\!\hat{m}_{0})q\!+\!
G\sum_{k=0}^{8}\big[(\bar{q}\lambda_{k}q)^{2}\!+\!
(\bar{q}i\gamma_{5}\lambda_{k}q)^{2}\big]\nonumber \\
           &&-K\big[\texttt{det}_{f}(\bar{q}(1+\gamma_{5})q)+\texttt{det}_{f}
(\bar{q}(1-\gamma_{5})q)\big]\nonumber \\ \nonumber \\
&&-U(\Phi[A],\bar{\Phi}[A],T),
\end{eqnarray}
where $q$ denotes the quark fields with three flavors, $u,\ d$, and
$s$; $\hat{m}_{0}=\texttt{diag}(m_{u},\ m_{d},\
m_{s})$ in flavor space; $G$ and $K$ are the four-point and
six-point interacting constants, respectively.  The $\hat{\mu}=diag(\mu_u,\mu_d,\mu_s)$ are the quark chemical potentials.

The covariant derivative in the Lagrangian is defined as $D_\mu=\partial_\mu-iA_\mu$.
The gluon background field $A_\mu=\delta_\mu^0A_0$ is supposed to be homogeneous
and static, with  $A_0=g\mathcal{A}_0^\alpha \frac{\lambda^\alpha}{2}$, where
$\frac{\lambda^\alpha}{2}$ is $SU(3)$ color generators.
The effective potential $U(\Phi[A],\bar{\Phi}[A],T)$ is expressed with the traced Polyakov loop
$\Phi=(\mathrm{Tr}_c L)/N_C$ and its conjugate
$\bar{\Phi}=(\mathrm{Tr}_c L^\dag)/N_C$. The Polyakov loop $L$  is a matrix in color space
\begin{equation}
   L(\vec{x})=\mathcal{P} exp\bigg[i\int_0^\beta d\tau A_4 (\vec{x},\tau)   \bigg],
\end{equation}
where $\beta=1/T$ is the inverse of temperature and $A_4=iA_0$.

The Polyakov-loop effective potential  is
%
\begin{eqnarray}
     \frac{U(\Phi,\bar{\Phi},T)}{T^4}&=&-\frac{a(T)}{2}\bar{\Phi}\Phi +b(T)\mathrm{ln}\big[1-6\bar{\Phi}\Phi\\ \nonumber
                                                &&+4(\bar{\Phi}^3+\Phi^3)-3(\bar{\Phi}\Phi)^2\big],
\end{eqnarray}
where
\begin{equation}
   \!a(T)\!=\!a_0\!+\!a_1\big(\frac{T_0}{T}\big)\!+\!a_2\big(\frac{T_0}{T}\big)^2 \,\,\,\texttt{and}\,\,\,\,\, b(T)\!=\!b_3\big(\frac{T_0}{T}\big)^3.
\end{equation}
The parameters $a_i$, $b_i$ listed in Table. \ref{tab:1} are fitted according to the lattice simulation of  QCD thermodynamics in
pure gauge sector. 
The $T_0=210$\, MeV
is implemented in the calculation.  
\begin{table}[ht]
\centering
\caption{Parameters in the Polyakov-loop potential~\cite{Robner07}}
\label{tab:1}
\begin{tabular*}{\columnwidth}{@{\extracolsep{\fill}}llll@{}}
\hline
\multicolumn{1}{@{}l}{$a_0$} & $a_1$ & $a_2$ & $b_3$\\
\hline
 $ 3.51$                   & -2.47        &  15.2      & -1.75               \\ 
 \hline
\end{tabular*}
\end{table}

The constituent quark mass in the mean field approximation can be derived as
\begin{equation}
M_{i}=m_{i}-4G\phi_i+2K\phi_j\phi_k\ \ \ \ \ \ (i\neq j\neq k),
\label{mass}
\end{equation}
 where $\phi_i$ stands for quark condensate of the flavor $i$.

The thermodynamical potential of bulk quark matter is derived as
\begin{eqnarray}
\Omega&=&U(\bar{\Phi}, \Phi, T)+2G\left({\phi_{u}}^{2}
+{\phi_{d}}^{2}+{\phi_{s}}^{2}\right)-4K\phi_{u}\,\phi_{d}\,\phi_{s} \nonumber\\
&&-2\int_\Lambda \frac{\mathrm{d}^{3}p}{(2\pi)^{3}}3(E_u+E_d+E_s) \nonumber \\
&&-2T \sum_{i=u,d,s}\int \frac{\mathrm{d}^{3}p}{(2\pi)^{3}} (\mathcal{Q}_1+\mathcal{Q}_2),
\end{eqnarray}
where $\mathcal{Q}_1=\mathrm{ln}(1
+3\Phi e^{-(E_i-\mu_i)/T}+3\bar{\Phi} e^{-2(E_i-\mu_i)/T}+e^{-3(E_i-\mu_i)/T})$,  $\mathcal{Q}_2=\mathrm{ln}(1+3\bar{\Phi} e^{-(E_i+\mu_i)/T}
+3\Phi e^{-2(E_i+\mu_i)/T}+e^{-3(E_i+\mu_i)/T}) $, and
 $E_i=\sqrt{\vec{p}^{\,2}+M_i^2}$ is the dispersion relation.  $\mu_i=\mu_B/3$ is taken for $u,d,s$ quark flavors. 
The pressure and energy density can be derived using the thermodynamic relations in the grand canonical ensemble as
\begin{equation}
\label{ }
p=-\Omega,\,\,\,\,\,\,\,\, \epsilon=-p+Ts+\mu_B \rho_B
\end{equation}

In the numerical calculation, a cut-off $\Lambda$ is implemented in 3-momentum
space for divergent integrations. We take the model parameters obtained in~\cite{Rehberg96}:
$\Lambda=602.3$ MeV, $G\Lambda^{2}=1.835$, $K\Lambda^{5}=12.36$,
$m_{u,d}=5.5$  and $m_{s}=140.7$ MeV, determined
by fitting $f_{\pi}=92.4$ MeV,  $M_{\pi}=135.0$ MeV, $m_{K}=497.7$ MeV and $m_{\eta}=957.8$ MeV. 

\section{Numerical results and discussions }

\begin{figure} [htbp]
\begin{minipage}{\columnwidth}
\centering
\includegraphics[scale=0.3]{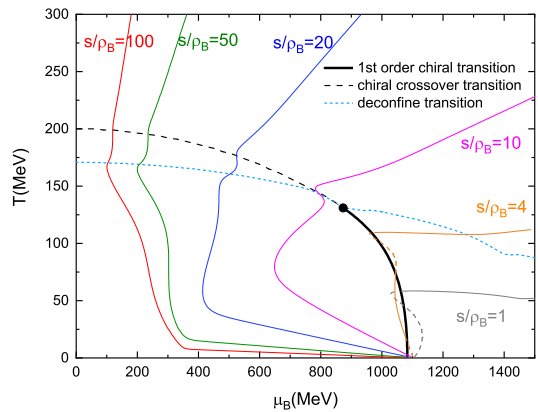}
\end{minipage}
\caption{  QCD phase diagram and the isentropic curves with  $s/\rho_B=100, 50, 20, 10, 4,1$  in $T-\mu_B$ plane.}
\label{fig:1}
\end{figure}

Firstly, we demonstrate in Fig.~\ref{fig:1} the isentropic curves with $s/\rho_B=100, 50, 20, 10, 4,1$ in the $T-\mu_B$ plane to
indicate the paths of fireball evolution for different collision energies. For $s/\rho_B=4$ and $1$, the traces in metastable and unstable region~(dashed parts) are also plotted in Fig.~\ref{fig:1}  to present the  full configuration. 
We also plot the chiral and deconfinement phase transition lines   in the PNJL quark model for  the convenience of latter discussion. We should note that the smooth crossover transformations take place in a wide range of temperature at small chemical potential. For more details of the phase transition, one can refer to \cite{Shao2018}.

\subsection {Speed of sound at constant $s/\rho_B$}

\begin{figure} [htbp]
\begin{minipage}{\columnwidth}
\centering
\includegraphics[scale=0.3]{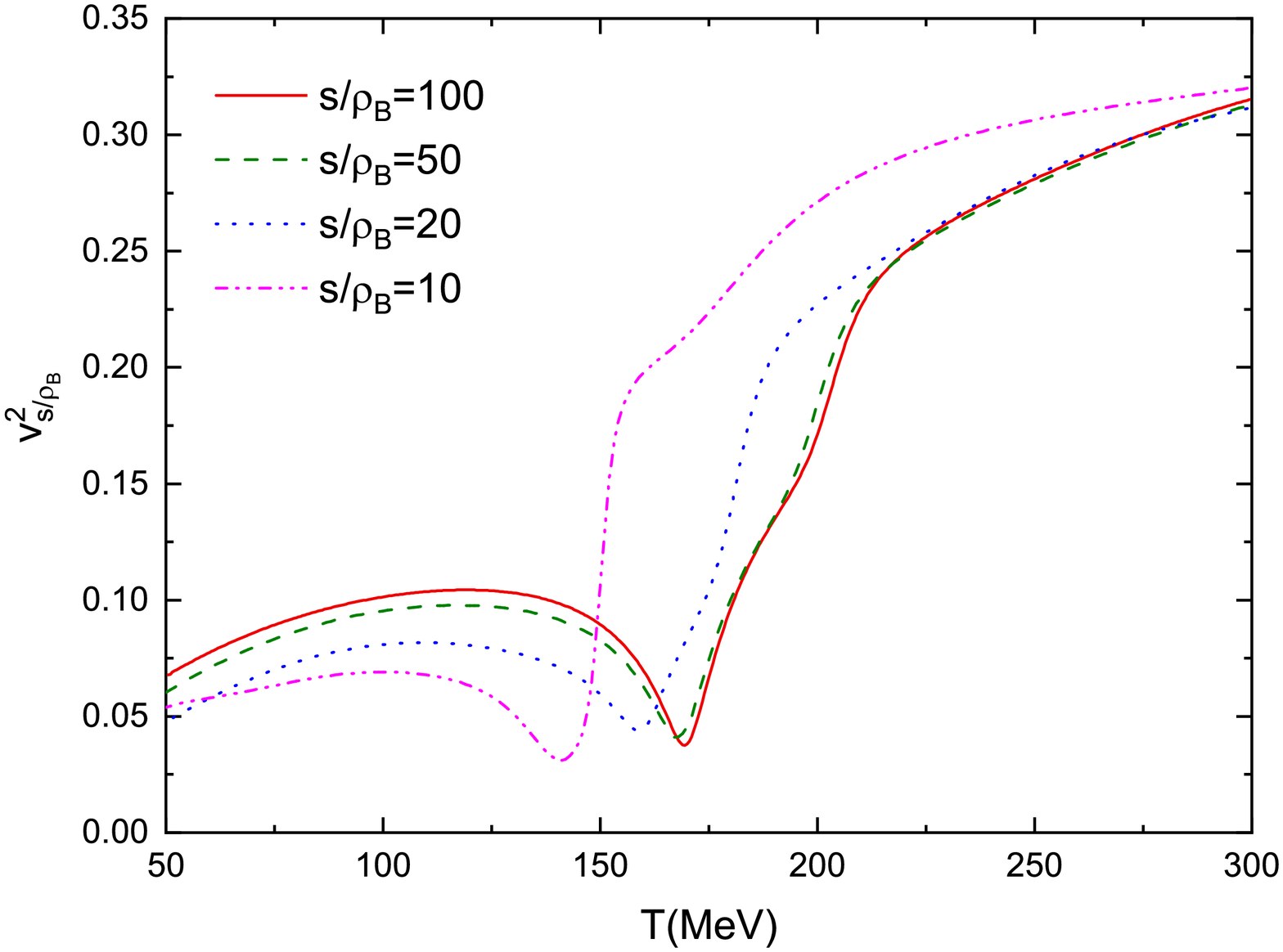}
\end{minipage}
\caption{  Squared speed of sound  as functions of temperature at  constant entropy density per baryon with $s/\rho_B=100, 50, 20, 10$, respectively.}
\label{fig:2}
\end{figure}

The value of $c^2_{s/\rho_B}$  at constant $s/\rho_B$ can reveal the speed of sound in the ideal hydrodynamic evolution.  
In Fig.~\ref{fig:2} , we plot the curves of $c^2_{s/\rho_B}$ as functions of temperature along the paths with  $s/\rho_B=100, 50, 20, 10,$ as shown in Fig.~\ref{fig:1}.  
It shows that, at the high-temperature side, $c^2_{s/\rho_B}$ approaches to $1/3$, the value of ideal gas. 
The $c^2_{s/\rho_B}$ decreases in the evolution with the decrease of temperature. A rapid decrease occurs in the chiral crossover region, and there exists a minimum for each $s/\rho_B$.

\begin{figure} [htbp]
\begin{minipage}{\columnwidth}
\centering
\includegraphics[scale=0.3]{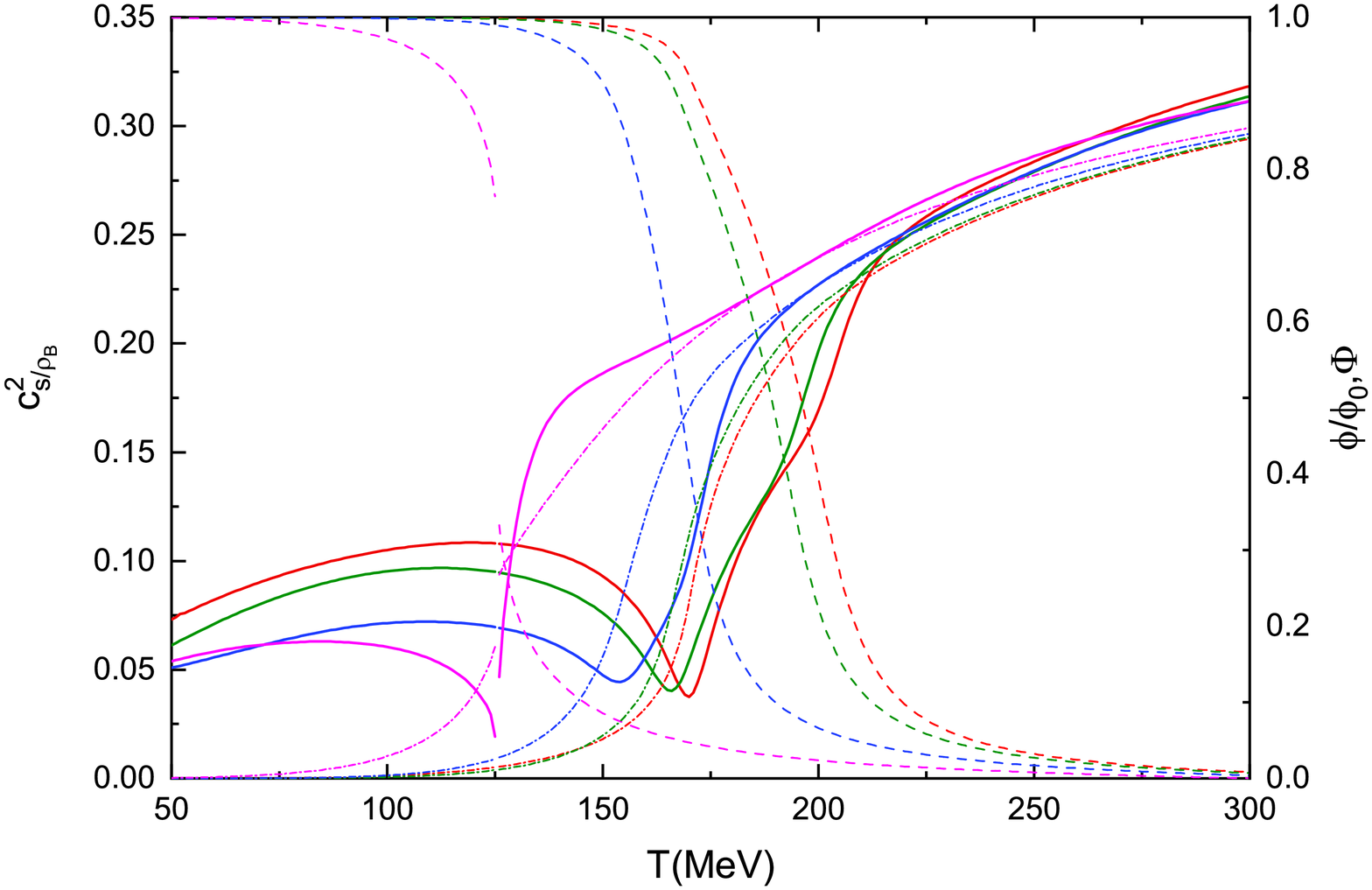}
\end{minipage}
\caption{ Values of $c^2_{s/\rho_B}$ (solid curves), scaled chiral condensate $\phi/\phi_0$ (dashed lines) and $\Phi$~(short dash dotted lines) as functions of temperature for $\mu_B=0, 300, 600, 900\,$ MeV with red, green, blue and magenta color, respectively.}
\label{fig:3}
\end{figure}

\begin{figure} [htbp]
\begin{minipage}{\columnwidth}
\centering
\includegraphics[scale=0.3]{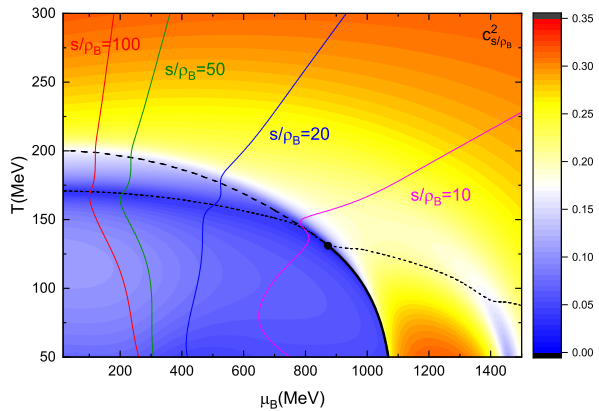}
\end{minipage}
\caption{ Contour map of $c^2_{s/\rho_B}$  in the $T-\mu_B$ plane. The same as in Fig~\ref{fig:1}, the curves with  $s/\rho_B=100, 50, 20, 10$ denote the paths of ideal hydrodynamic evolution in $T-\mu_B$ plane}
\label{fig:4}
\end{figure}

To indicate the relation between the speed of sound and the chiral and deconfinement phase transitions,  we plot in Fig.~\ref{fig:3} the values of $c^2_{s/\rho_B}$, the scaled chiral condensate $\phi/\phi_0$ of $u(d)$ quark, and $\Phi$ as functions of temperature with four given baryon chemical potentials $\mu_B=0, 300,600,900\,$MeV. 
We also present in Fig.~\ref{fig:4} the full contour map of $c^2_{s/\rho_B}$  in the $T-\mu_B$ plane. 

Fig.~\ref{fig:3} and Fig.~\ref{fig:4}  indicate that $c^2_{s/\rho_B}$ decreases rapidly in the crossover region with the increase of chiral condensate~(dynamical quark mass). It means that the masses of quasiparticles play a crucial role on the speed of sound in QCD matter. A minimum of $c^2_{s/\rho_B}$ appears in the low-temperature side of the chiral crossover transformation, which happens to coincide with the deconfinement phase transition. Some similar features were discussed in \cite{Motta18}. The dip structure shown in Fig.~\ref{fig:2} and Fig.~\ref{fig:4} was also discovered in the LQCD simulation at zero chemical potential~\cite{Aoki06, Borsanyi14, Bazavov14, Philipsen13, Borsanyi20}. However, such a feature does not appear in the NJL model which lacks the confinement, at least not obvious~\cite{Ghosh06,Marty13,Deb16,Saha18}. It hints that the deconfinement transformation involved simultaneously also plays a vital role in the evolution.

\begin{figure} [htbp]
\begin{minipage}{\columnwidth}
\centering
\includegraphics[scale=0.3]{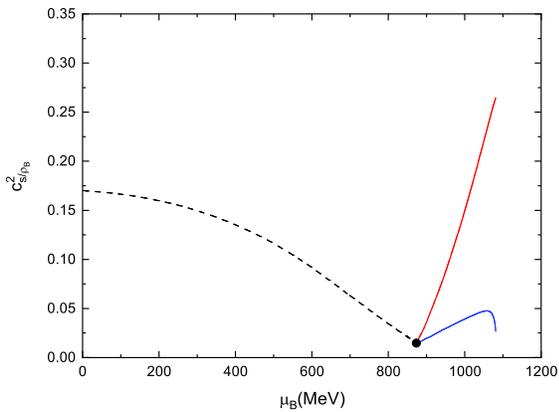}
\end{minipage}
\caption{ Values of $c^2_{s/\rho_B}$ along the chiral phase transition line. The dashed line is the result of chiral crossover transition. There are two branches for the first-order transition. One~(red line) is for the chiral restored phase, the other one~(blue line) is for the chiral breaking phase.}
\label{fig:5}
\end{figure}

In Fig.~\ref{fig:5}, we present the value of $c^2_{s/\rho_B}$ along the chiral phase transition line. It indicates that $c^2_{s/\rho_B}$ decreases along the chiral crossover transition line from zero chemical potential to the CEP. A noticeable feature is that  $c^2_{s/\rho_B}$ approaches to almost zero at the CEP. This is consistent with the critical slowing down, since in the critical region a small perturbation can induce drastic density fluctuation, to the disadvantage of  the spread of wave.   
There are two branches of $c^2_{s/\rho_B}$ for the first-order phase transition. One branch is for the chiral restored phase, the other one is for the chiral breaking phase. It is obvious that  $c^2_{s/\rho_B}$ in the chiral restored phase is much larger than that in the chiral breaking phase. This reflects again that the dynamical quark mass affects greatly on the EOS of quark matter.

Fig.~\ref{fig:4} also shows that there exist an area in the quarkyonic phase where $c^2_{s/\rho_B}$ has relatively larger values. It is not difficult to understand such a result from the restoration of chiral phase transition of $u, d$ quark. However, strange quark still has a larger mass.
With the increase of chemical potential, the value of $c^2_{s/\rho_B}$ decreases again  before the occurrence of the first-order phase transition of strange quark. 
At extreme high baryon chemical potential, $c^2_{s/\rho_B}$ increase again after the chiral restoration of strange quark. It means that $c^2_{s/\rho_B}$ at low temperature has a hierarchy  along the chemical potential since the chiral phase transitions for $u(d)$ quark and $s$ quark take place in sequence.

\subsection {Speed of sound at constant baryon density and entropy density}
In the following we study the squared speed of sound at constant baryon density and constant entropy density. The two definitions of sound velocity are usually used in the intermediate description of hydrodynamic evolution.
For example, the temporal derivatives of temperature and  chemical potential are functions of $c_s^2$ and $c_{\rho_B}^2$, as
\begin{equation}
\label{ }
\partial_0 T=-c_{\rho_B}^2  T\, \bold{\nabla}\cdot \bold{u},    
\end{equation}
and 
\begin{equation}
\label{ }
\partial_0 \mu_B=-c_{s}^2  \mu_B\, \bold{\nabla}\cdot \bold{u},    
\end{equation}
where $\bold u$ denotes the space component of four-velocity.

\begin{figure} [htbp]
\begin{minipage}{\columnwidth}
\centering
\includegraphics[scale=0.3]{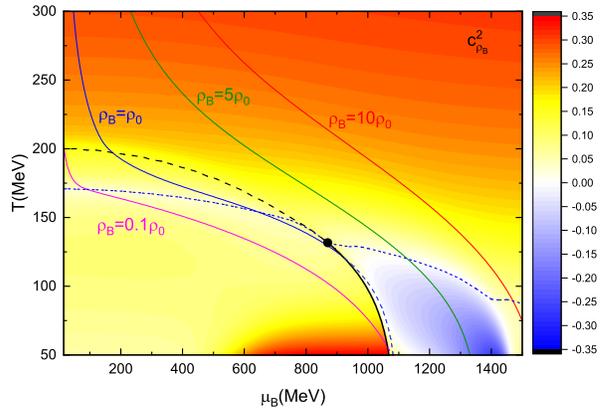}
\end{minipage}
\caption{   Contour map of $c^2_{\rho_B}$  in the $T-\mu_B$ plane. Four curves with $\rho_B=0.1, 1, 5, 10 \rho_0$ are also plotted to show the relations between baryon density and $(T, \mu_B)$.}
\label{fig:6}

\end{figure}

In Fig.~\ref{fig:6}, we show  the contour map of $c^2_{\rho_B}$ in the $T-\mu_B$ plane. To demonstrate the path of constant baryon density,  four curves with $\rho_B=0.1, 1, 5, 10 \rho_0$~($\rho_0$ is the saturation density of nuclear matter) are  plotted.
The contour plot shows that the squared speed of sound $c^2_{\rho_B}$ also approaches to $1/3$ at high temperature.  Similar to the behavior of $c^2_{s/\rho_B}$,
a minimum of $c^2_{\rho_B}$ appears along the deconfinement phase transition in the crossover region. 

However, the deference between $c^2_{\rho_B}$ and $c^2_{s/\rho_B}$ grows with the decrease of temperature. The paths of constant $\rho_B$ in $T-\mu_B$ plane are almost perpendicular with those of constant $s/\rho_B$ in a wide area. It just indicates the sensitivity of the change of interaction to temperature and chemical potential.
It is interesting that there exists an area of negative $c^2_{\rho_B}$ in the quarkyonic phase. In this region, with the increase of energy density, the pressure decreases along the path of a constant density. 

\begin{figure} [htbp]
\begin{minipage}{\columnwidth}
\centering
\includegraphics[scale=0.3]{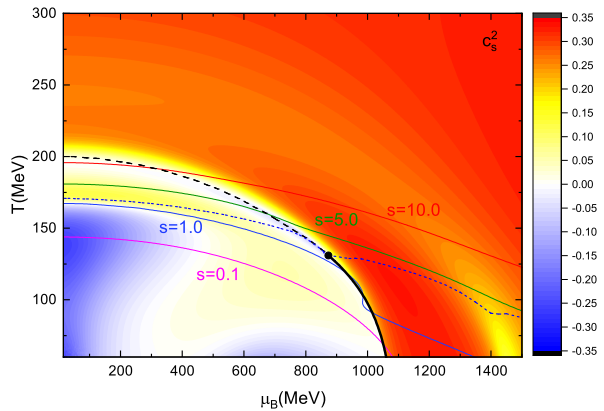}
\end{minipage}
\caption{ Contour map of $c^2_{s}$  in the $T-\mu_B$ plane. Four curves with $s=0.01, 1, 5, 10 fm^{-3}$ to show the relation between entropy density and $(T, \mu_B)$.}
\label{fig:7}
\end{figure}

We present the contour map of  $c_{s}^2$ in Fig.~\ref{fig:7}. Similar to Fig.~\ref{fig:4} and Fig.~\ref{fig:6}, we plot four curves with $s=0.01, 1, 5, 10 fm^{-3}$ to show the relation between entropy density and $(T, \mu_B)$.  The numerical result also shows that $c^2_{s}$  is close to $1/3$  at high temperature. Besides, there is a large area where the value of $c^2_{s}$ is negative at the lower left  in figure \ref{fig:7}. The other negative area is in a narrow range near the chiral crossover phase transition not far from the CEP. 


As shown in Fig.~\ref{fig:6} and Fig.~\ref{fig:7},  the negative values appear for both $c^2_{\rho_B}$ and  $c_{s}^2$. More accurately, the partial derivative of $(\frac{\partial p}{\partial \epsilon})_{\rho_{B}}$ or $(\frac{\partial p}{\partial \epsilon})_s$ is negative in some regions.
It is not worrying about the negative values, because  $c^2_{s/\rho_B}$, rather than $c^2_{\rho_B}$ and  $c_{s}^2$, is the sound velocity under the adiabatic condition in the hydrodynamic evolution.  Nevertheless, it is very significative to find the underlying reason to explain the behaviors of $c^2_{\rho_B}$ and  $c_{s}^2$ in the full phase diagram.

Applying the thermodynamic relation
\begin{equation}
\label{ }
\bigg(\frac{\partial \mu_B}{\partial T}\bigg)_{s/\rho_B}\cdot \bigg(\frac{\partial T}{\partial {(s/\rho_B)}}\bigg)_{\mu_B}\cdot\bigg(\frac{\partial (s/\rho_B)}{\partial {\mu_B}}\bigg)_{T}=-1,
\end{equation}
and the formulae of Eqs.~(\ref{crho}) and (\ref{cs}), we can obtain 
\begin{equation}
\label{ }
\bigg(\frac{\partial \mu_B}{\partial T}\bigg)_{s/\rho_B}=\frac{\mu_B}{T}\frac{c^2_s}{c^2_{\rho_B}}=\frac{\mu_B}{T}\frac{(\frac{\partial p}{\partial \epsilon})_{s}}{(\frac{\partial p}{\partial \epsilon})_{\rho_{B}}}.
\end{equation}
This equation indicates clearly that $(\frac{\partial p}{\partial \epsilon})_{s}$  and $(\frac{\partial p}{\partial \epsilon})_{\rho_{B}}$    have opposite sign for the case of $\big(\frac{\partial \mu_B}{\partial T}\big)_{s/\rho_B}<0$. 

As shown by the isentropic curves in Fig.~\ref{fig:1}, such a situation does exist at the lower left of the $T-\mu_B$ phase diagram and a small neighboring area of chiral-crossover transformation near the CEP, as well as a part of the quarkyonic phase. The contour maps in Fig.~\ref{fig:6} and Fig.~\ref{fig:7} precisely proves this conclusion.
It is instructive to take this result in hydrodynamics simulation when the critical region and low-temperature region is involved. 

In the following, we try to explore the physical meaning of the negative values of $(\frac{\partial p}{\partial \epsilon})_{\rho_{B}}$ or $(\frac{\partial p}{\partial \epsilon})_s$ in quark matter. Usually, a disturbance  travels through matter with the sound velocity since it obeys the propagation equation of sound wave. To build the sound wave equation,   $\left(\frac{\partial p}{\partial \epsilon}\right)_{x}>0$ is required to make sure  that the disturbance (pressure gradient) can propagate in waves. However,  the non-perturbative interaction of QCD is much more complicated than ordinary matter, and there exists some regions where $c_{s}^{2}=\left(\frac{\partial p}{\partial \epsilon}\right)_{s}<0$ or $c_{\rho_{B}}^{2}=\left(\frac{\partial p}{\partial \epsilon}\right)_{\rho_{B}}<0$ under the condition of $\left(\frac{\partial \mu_B}{\partial T}\right)_{s/\rho_B}<0$. 
 In this case,  the wave equation  at constant  $s$ or $\rho_B$  will be broken, and turns into another equation whose solution is a decay function. It means the disturbance at constant $s$ or $\rho_B$ will not propagate in the form of sound wave, but decay soon. The main reason is that the increase of pressure is associated with the decrease of energy density, contrary to the sound wave propagation condition.  
Therefore, the negative values of $c_{s}^{2}=\left(\frac{\partial p}{\partial \epsilon}\right)_{s}$ or $c_{\rho_{B}}^{2}=\left(\frac{\partial p}{\partial \epsilon}\right)_{\rho_{B}}$ in Fig.~\ref{fig:6} and Fig.~\ref{fig:7}  indicate the region where the pressure gradient can not be spread out in waves but decay under the condition of constant $s$ or $\rho_B$, respectively.  Correspondingly,  $c_s$ and $c_{\rho_{B}}$ in such a situation are no longer the real sound velocities in the conventional sense.

We should also note that, for the adiabatic sound velocity, $c_{s/\rho_{B}}^{2}=\left(\frac{\partial p}{\partial \epsilon}\right)_{s/\rho_{B}}$  is always positive in the stable phase, but negative values possibly appear in the unstable or metastable phase when the spinodal structure of the first-order phase transition is included~(It is outside the scope of this work). A further study including the metastable and unstable phase  associated with the first-order phase transition will be done in the future to provide more insights on sound velocity in QCD matter.

\subsection {Speed of sound at constant temperature and chemical potential}

\begin{figure} [htbp]
\begin{minipage}{\columnwidth}
\centering
\includegraphics[scale=0.3]{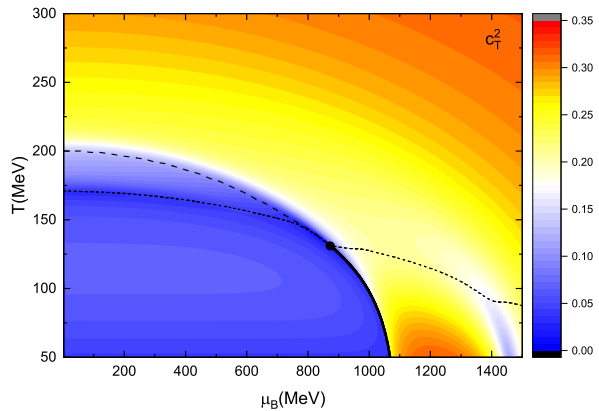}
\end{minipage}
\caption{ Contour map of $c^2_{T}$  in the $T-\mu_B$ plane.}
\label{fig:8}
\end{figure}

\begin{figure} [htbp]
\begin{minipage}{\columnwidth}
\centering
\includegraphics[scale=0.3]{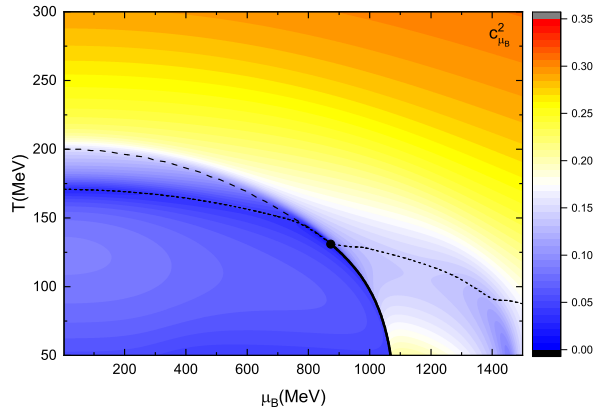}
\end{minipage}
\caption{ Contour map of $c^2_{\mu_B}$  in the $T-\mu_B$ plane.}
\label{fig:9}
\end{figure}

The contour maps of $c^2_T$ at constant temperature and $c^2_{\mu_B}$ at constant chemical potential are demonstrated in Fig. \ref{fig:8} and Fig. \ref{fig:9}, respectively.  The two figures show that the squared speeds of sound  $c^2_T$ and $c^2_{\mu_B}$ are both close to $1/3$ at high temperature.
The contour of $c^2_{\mu_B}$ at smaller chemical potential is similar to $c^2_{s/\rho_B}$ since the paths of constant $s/\rho_B$ are nearly parallel to the temperature axis. Correspondingly, the contour of $c^2_{T}$ at larger chemical potential and low temperature is similar to that of $c^2_{s/\rho_B}$, because  the curves of constant $s/\rho_B$ in this region are almost parallel to the chemical potential axis. 

\section{Summary}   
In this work, we studied the squared speed of sound in QCD matter under five different definitions in the PNJL model using the grand canonical ensemble.
The different definitions of speed of sound  mean that the derivative of $p$ respect to $\epsilon$ are taken along different orientations, which can indicate important properties of the equation of state. The squared speed of sound under five definitions are all approaching to 1/3, the Stefan-Boltzmann limit, at high temperature. It indicates that the speed of sound of quark-gluon plasma at high temperature is not sensitive to the path of taking the derivative of $p$ respect to $\epsilon$. It is just the feature of the chiral restored and deconfined phase.

The situations become complicated at  finite temperature and chemical potential due to the non-perturbative interaction and ($T$, $\mu_B$) dependent phase transition. we analyzed the relations between the squared speed of sound and the phase transitions.  From the perspective of fluid evolution, the value of $c^2_{s/\rho_B}$ has a rapid decrease in the chiral crossover region from higher temperature to lower temperature. A minimum appears at the low-temperature side of the chiral crossover transformation, which coincides with the deconfinement phase transition in the PNJL model.   The value of $c^2_{s/\rho_B}$  approaches to almost zero at the CEP. There also exists a hierarchical behavior of $c^2_{s/\rho_B}$ between $u(d)$ and $s$ quark at low temperature with the increasing chemical potential.

We also discussed the relations of squared speed of sound under different definitions, in particular the correlation between $c^2_s$ and $c^2_{\rho_B}$. We found  $(\frac{\partial p}{\partial \epsilon})_{\rho_{B}}$ and $(\frac{\partial p}{\partial \epsilon})_s$  take opposite sign for the case of $\big(\frac{\partial \mu_B}{\partial T}\big)_{s/\rho_B}<0$. Such a situation indeed exists in the QCD phase diagram. 
 For the case of $\left(\frac{\partial p}{\partial \epsilon}\right)_{s}<0$ or $\left(\frac{\partial p}{\partial \epsilon}\right)_{\rho_{B}}<0$, $c_s$ and $c_{\rho_{B}}$ in such a situation are no longer the real sound velocities. 
Attentions should be paid when the corresponding speed of sound is taken in hydrodynamics simulation in the critical region and the low-temperature region involved. 

\begin{acknowledgements} 
This work is supported by the National Natural Science Foundation of China under
Grant No. 11875213.
\end{acknowledgements}

\end{document}